\documentclass[aps,prc,twoside,twocolumn,nofootinbib,10pt,superscriptaddress,%
showpacs,floatfix]{revtex4-1}

\usepackage{amsmath,amssymb,bm}
\usepackage{graphicx}
\usepackage{slashed}

\usepackage{tensor}

\allowdisplaybreaks

\begin{document}

\title{\boldmath Investigation of the $nn\Lambda$ bound state in pionless effective theory}

\author{Shung-Ichi Ando}
\email{sando@sunmoon.ac.kr}
\affiliation{Department of Information Communication \& Display Engineering, 
Sunmoon University, Asan, Chungnam 31460, Korea}

\author{Udit Raha}
\email{udit.raha@iitg.ernet.in}
\affiliation{Department of Physics, Indian Institute of Technology Guwahati, 
781 039 Assam, India}

\author{Yongseok Oh}
\email{yohphy@knu.ac.kr}
\affiliation{Department of Physics, Kyungpook National University, 
Daegu 41566, Korea}
\affiliation{Asia Pacific Center for Theoretical Physics, Pohang, 
Gyeongbuk 37673, Korea}

\date{}

\begin{abstract}
The possibility of an $nn\Lambda$ bound state is investigated in the framework of pionless effective 
field theory at leading order. 
A system of coupled integral equations are constructed in the spin-isospin basis, of which numerical 
solutions are investigated.
In particular, we make use of the limit cycle behavior, i.e., cyclic singularities of coupled integral 
equations of the system, which would be associated with the formation of a three-body bound state, 
so-called the Efimov state, in the unitary limit.
Furthermore, we find that, when the sharp momentum cutoff introduced in the integral equations is taken 
significantly larger than the hard scale of the effective theory, the coupling of a three-body contact 
interaction becomes cyclically singular indicating the onset of Efimov-like bound state formation. 
However, the paucity of empirical information to determine the parameters of the theory precludes a 
definitive conclusion on the existence of such a bound state. 
As a simple test of the feasibility of the $nn\Lambda$ bound system in nature, we explore the 
cutoff dependence of the theory, and uncertainties of the present study are discussed as well.
\end{abstract}

\pacs{
11.10.Hi, 
13.75.Ev, 
21.45.-v, 
21.80.+a. 
}

\maketitle


\section{Introduction}

Recently the formation of \nuclide[3][\Lambda]{n}, the bound state of two neutrons and one 
$\Lambda$ hyperon ($nn\Lambda$ system), has been suggested by the experimental investigation 
of the HypHI Collaboration~\cite{HypHI-13}.%
\footnote{The possibility of such an $nn\Lambda$ bound state was also suggested by 
the lattice QCD simulation of Ref.~\cite{NPLQCD-12} in the limit of flavor SU(3) symmetry.}
If confirmed, this observation would be a crucial indication of the discovery of a new exotic bound 
state, namely, a nucleus without a proton. 
Subsequent theoretical studies~\cite{GV14,HOGR14,GG14,RWZ14}, however, questioned the claim of 
Ref.~\cite{HypHI-13} because of the inconsistencies of the putative $nn\Lambda$ bound state with 
other observables such as the $N\Lambda$ scattering data, hypertriton binding energy, and the 
energy gaps between the ground and first excited states of \nuclide[4][\Lambda]{H} and 
\nuclide[4][\Lambda]{He}. 
Such theoretical works were mostly based on the estimations employing standard potential model 
approaches, which include the $\Lambda N $-$\Sigma N$ mixing that play a crucial role in 
charge symmetry breaking effects and determination of the spin-parity quantum numbers of the 
ground state.
In fact, the seminal theoretical work of Ref.~\cite{DD59} already reported a long time ago 
on the non-existence of $nn\Lambda$ bound state where model parameters of a variational 
calculation were tuned by the hypertriton binding energy.%
\footnote{
Other earlier theoretical studies on this system can be found, e.g., in Refs.~\cite{MKGS95,BRS07}.}

In the present exploratory study, we investigate the possibility of a bound $nn\Lambda$ system 
employing a pionless effective field theory (EFT) at leading order (LO) following the approach of 
Ref.~\cite{Hammer01}, which was used to investigate the $pn\Lambda$ system and the hypertriton 
bound state. 
Previously, low-energy EFTs were constructed by two of the present authors~\cite{AYO13,AO14} for 
investigating the hypernuclei \nuclide[4][\Lambda\Lambda]{H} and \nuclide[6][\Lambda\Lambda]{He} 
as possible $\Lambda\Lambda d$ and $\Lambda\Lambda\alpha$ bound states, respectively. 
These investigations revealed substantial indications, corroborating previous claims, that 
\nuclide[4][\Lambda\Lambda]{H} is likely to form  a bound state and the bound state of 
\nuclide[6][\Lambda\Lambda]{He} could be an Efimov state.

A low-energy EFT is constructed by introducing a hard scale $\Lambda_H$ that separates the 
relevant low-energy degrees of freedom from the irrelevant high-energy degrees of freedom which 
are to be ``integrated out.''
The advantage of this approach is that it provides with a model-independent and systematic 
calculational technique with a small number of coupling constants that embody all the ignorance 
about the short-distance dynamics. 
For a review on the details of this subject we refer the reader to Refs.~\cite{BV02c,BH04} and 
references therein.

Throughout this work, we are dealing with a three-body system which, if bound, is likely to have 
a binding energy much smaller than the pion mass.%
\footnote{The estimated binding energy of the putative $nn\Lambda$ bound state in 
Ref.~\cite{HypHI-13} is at most a few MeV.}
Therefore, we can choose the pion mass as the hard scale, i.e., $\Lambda_H \sim m_\pi^{}$, 
so that the pions are integrated out and are not explicitly introduced in the theory. 
We may additionally regard the mass difference between the $\Lambda$ and $\Sigma$ hyperons, 
i.e., $\delta m = m_\Sigma^{} - m_{\Lambda}^{} \simeq 80~\mbox{MeV}$, as of the same order 
as the hard scale $\Lambda_H$. 
Then our effective Lagrangian can be written explicitly in terms of the neutron and $\Lambda$ 
fields along with their interactions described by the contact terms, which will be determined in a  
phenomenological way.

In addition, we will make use of the cyclic singularities that arise in the solutions for the coupled 
integral equations in the asymptotic limit~\cite{Danilov61}. 
Such singularities are renormalized by introducing a suitably large momentum cutoff $\Lambda_c$ 
($\Lambda_c \gtrsim \Lambda_H$) in the loop integrations at the cost of introducing three-body 
counter terms at LO. 
Consequently, in order to absorb this cutoff dependence, the corresponding three-body coupling 
may exhibit a cyclic renormalization group (RG) evolution termed as the limit cycle~\cite{Wilson71}. 
The cyclic singularities are associated with the occurrence of bound states, known as the Efimov 
states, in the resonant/unitary limit~\cite{Efimov71}. 
In our analysis, we vary the magnitude of the cutoff within a reasonable range to investigate its 
sensitivity to the formation of bound states. 
This is a simple test one can perform for checking the feasibility of the $nn\Lambda$ system 
as a three-body bound system. 
The main purpose of the present investigation is, thus, to explore the putative bound state of 
the $nn\Lambda$ system in the context of a modern EFT method mentioned above.

This paper is organized as follows.
In Sec.~\ref{sec:Lag}, the effective Lagrangian at LO for the system is introduced. 
Then the renormalized dressed two-body propagators are defined and the three-body coupled integral 
equations are derived in Sec.~\ref{sec:amp}.
Before numerically solving the coupled integral equations, we obtain an analytical expression of 
a scale-invariant equation needed to examine the limit cycle behavior in the asymptotic limit by 
assuming that the neutron and $\Lambda$ hyperon have the same mass.
Section~\ref{sec:results} is devoted to numerical solutions of the coupled integral equations for the 
$nn\Lambda$ system with physical baryon masses, and we test if the three-body contact interaction 
exhibits cyclic singularities, typically associated with the Efimov states, even at low and intermediate 
momenta being away from the unitary limit. 
We further investigate the possibility of bound state formation in the absence of the contact interaction, 
when the sharp cutoff $\Lambda_c$ is chosen to be significantly larger than $\Lambda_H$. 
Section~\ref{sec:dis} summarizes the present work with possible implications of our results and the 
uncertainties involved in the present approach.


\section{Effective Lagrangian}
\label{sec:Lag}

The relevant non-relativistic effective Lagrangian for the $nn\Lambda$ system at LO consistent with 
parity, charge conjugation symmetry, time-reversal invariance, and small-velocity Lorentz 
transformation reads
\begin{equation}
\mathcal{L} = \mathcal{L}_n^{} + \mathcal{L}_\Lambda^{} + \mathcal{L}_{s(nn)}^{}
+ \mathcal{L}_{s(n\Lambda)}^{} + \mathcal{L}_{t(n\Lambda)}^{}
+ \mathcal{L}_{\mbox{\scriptsize 3-body}} ,
\label{eq:Lag0}
\end{equation}
where the elementary fields of our EFT are the neutron field $\mathcal{B}_n$ and the $\Lambda$ 
hyperon field $\mathcal{B}_\Lambda$ of which one-body Lagrangian are represented by 
$\mathcal{L}_n^{}$ and $\mathcal{L}_\Lambda^{}$, respectively.
The $S$-wave two-body Lagrangian for the spin singlet $nn$ channel, spin singlet $n\Lambda$ 
channel, and spin triplet $n\Lambda$ channel are respectively represented by $\mathcal{L}_{s(nn)}$, 
$\mathcal{L}_{s(n\Lambda)}$, and $\mathcal{L}_{t(n\Lambda)}$.
The composite dibaryon fields are introduced and denoted by $s_{(nn)}$ for the spin singlet $nn$, 
and $s_{(n\Lambda)}$ and $t_{(n\Lambda)}$ for the spin singlet and spin triplet $n\Lambda$ 
systems, respectively.
The three-body interaction Lagrangian is represented by $\mathcal{L}_{\mbox{\scriptsize 3-body}}$.

The one-body Lagrangian $\mathcal{L}_n^{}$ and $\mathcal{L}_\Lambda^{}$ of Eq.~(\ref{eq:Lag0})
are given as~\cite{BKM95,AM97} 
\begin{eqnarray}
\mathcal{L}_n &=& \mathcal{B}_n^\dagger 
\left[ i v \cdot \partial  +\frac{(v \cdot \partial)^2 - \partial^2}{2m_n^{}}
\right] \mathcal{B}_n + \cdots  ,
\\[2mm]
\mathcal{L}_\Lambda &=& \mathcal{B}_\Lambda^\dagger 
\left[ i v \cdot \partial +\frac{(v \cdot \partial)^2 - \partial^2}{2m_\Lambda^{}}
\right] \mathcal{B}_\Lambda + \cdots ,
\end{eqnarray}
where $v^\mu$ is a velocity four-vector chosen as $v^\mu=(1, \bm{0})$.
The neutron and $\Lambda$ hyperon masses are given by $m_n^{}$ and $m_\Lambda^{}$, 
respectively.
The ellipses denote higher order terms, which are not required at the accuracy of the present analysis.

The $S$-wave two-body interactions are written in terms of the dibaryon fields 
as~\cite{BS00,AH04,AK06b}
\begin{widetext}
\begin{eqnarray}
\mathcal{L}_{s(nn)} &=& \sigma_{s(nn)}^{} \, s_{(nn)}^\dagger 
\left[ i v \cdot \partial + \frac{(v \cdot \partial)^2 - \partial^2}{4m_n^{}} 
+ \Delta_{s(nn)} \right] s_{(nn)}^{}
- y_{s(nn)} \left[ s_{(nn)}^\dagger \left( \mathcal{B}_n^T P_{(nn)}^{({}^1S_0)} 
\mathcal{B}_n \right) + \mbox{ H.c.} \right] + \cdots ,
\\
\mathcal{L}_{s(n\Lambda)} &=& \sigma_{s(n\Lambda)}^{} \, s_{(n\Lambda)}^\dagger \left[
i v\cdot \partial + \frac{(v \cdot \partial)^2 - \partial^2}{2(m_n^{} + m_\Lambda^{})}
+ \Delta_{s(n\Lambda)} \right] s_{(n\Lambda)}^{}
- y_{s(n\Lambda)} \left[ s_{(n\Lambda)}^\dagger \left( \mathcal{B}_n^T P_{(n\Lambda)}^{(^1S_0)} 
\mathcal{B}_\Lambda \right) + \mbox{ H.c.} \right] + \cdots ,
\\
\mathcal{L}_{t(n\Lambda)} &=& 
\sigma_{t(n\Lambda)}^{} \, t^\dagger_{(n\Lambda)k} \left[ i v\cdot \partial 
+ \frac{(v \cdot \partial)^2 - \partial^2}{2(m_n^{} + m_\Lambda^{})} 
+ \Delta_{t(n\Lambda)} \right] t_{(n\Lambda)k}^{}
- y_{t(n\Lambda)} \left[ t_{(n\Lambda)k}^\dagger \left( \mathcal{B}_n^T P_{(n\Lambda)k}^{(^3S_1)} 
\mathcal{B}_\Lambda \right) 
+ \mbox{ H.c.} \right] + \cdots ,
\end{eqnarray}
\end{widetext}
where $\sigma_{s(nn)}^{}$, $\sigma_{s(n\Lambda)}^{}$, and $\sigma_{t(n\Lambda)}^{}$ are sign 
factors, and $\Delta_{s(nn)}$, $\Delta_{s(n\Lambda)}$, and $\Delta_{t(n\Lambda)}$ are the respective 
mass differences between the corresponding dibaryon and its constituent elementary particles. 
The coupling constants are denoted by $y_{s(nn)}^{}$, $y_{s(n\Lambda)}^{}$, and 
$y_{t(n\Lambda)}^{}$, which will be determined by the $S$-wave effective range parameters such 
as the scattering lengths and effective ranges~\cite{BS00,AH04,AK06b} as will be discussed in the 
next section.
The spin projection operators introduced in the above Lagrangian are defined as
\begin{eqnarray}
&& 
P_{(nn)}^{(^1S_0)} = - \frac{i}{2} \sigma_2^{} \,,
\quad
P_{(n\Lambda)}^{(^1S_0)} = - \frac{i}{\sqrt2} \sigma_2^{} \,,
\nonumber\\  &&
P_{(n\Lambda)k}^{(^3S_1)} = - \frac{i}{\sqrt2} \sigma_2^{} \, \sigma_k^{} \, ,
\end{eqnarray}
where the difference in the factors in the definitions of the spin singlet projection operators, 
$P^{(^1S_0)}_{(nn)}$ and $P^{(^1S_0)}_{(n\Lambda)}$, arises from the existence of two identical 
particles in the $nn$ channel.

As will be demonstrated in the next section, the solutions for the coupled integral equations in the 
spin doublet channel of the $nn\Lambda$ system exhibit cyclic singularities in the asymptotic limit. 
To renormalize the cyclic singularities, it mandates the inclusion of three-body counter terms 
already at LO, where the singularities associated with the short range part of the 
one-neutron/$\Lambda$-exchange interactions are needed to be canceled out. 
In three-nucleon systems, the general expression for three-nucleon counter term is constructed in 
the Wigner SU(4)-symmetric limit~\cite{Griess04,Griess05}. 
Moreover, in the case of the hypertriton, i.e., the spin doublet $np\Lambda$ channel, the general 
expression for the three-body counter term is obtained in the limit where the mass difference 
between the nucleon and the $\Lambda$ hyperon is ignored~\cite{Hammer01}. 
Following this approach, as will be shown in the next section, one may construct an analogous 
expression for the counter term using the projection operator for the specific \textit{diagonal\/} 
mode required to renormalize the asymptotic cyclic singular behavior in the same mass limit. 
However, with the physical value of the mass difference $m_\Lambda^{} - m_n^{}$ taken into 
account, a more systematic way of analytical determination of the counter term is challenging. 
Furthermore, we will make use of the RG evolution of the bound states that exhibits the periodic 
occurrence of critical cutoff values, corresponding to the vanishing points of the three-body contact 
interaction, in studying the limit cycle behavior of this system at threshold, i.e., with zero three-body 
binding energy.

In our present investigation, therefore, we introduce the three-body counter-term Lagrangian for the 
spin-doublet channel as 
\begin{eqnarray}
\mathcal{L}_{\mbox{\scriptsize 3-body}} &=&
- \frac16 m_\Lambda^{} y_{t(n\Lambda)}^2 
\frac{g(\Lambda_c)}{\Lambda_c^2}  t^\dagger_{(n\Lambda)i} \mathcal{B}_n^\dagger
\sigma_i^{} \sigma_j^{} \mathcal{B}_n t_{(n\Lambda)j}^{} + \dots \, ,
\nonumber \\
\label{eq;3bci}
\end{eqnarray}
where $g(\Lambda_c)$ is a cutoff-dependent coupling constant.
In general, the 3-body contact interaction of Eq.~(\ref{eq;3bci}) contains other terms that involve the 
$nn$ dibaryon and $\Lambda$ hyperon fields as well as the spin singlet $n\Lambda$ dibaryon and 
neutron fields with different coupling strengths. 
However, as will be discussed later, the paucity of data on the $nn\Lambda$ system at the present 
stage does not allow the estimation of these couplings. 
Thus, the term shown in Eq.~(\ref{eq;3bci}) will adequately serve our purpose in this explorative 
study where we hope to capture certain universal features of such bound systems even without 
resorting to a more sophisticated EFT analysis.
In principle, the above coefficient $g(\Lambda_c)$ should be renormalized as a function of the same 
cutoff parameter $\Lambda_c$ that is introduced in the coupled integral equations for the three-body 
system and thereby fixed by the experimental/empirical data on a three-body observable of the 
$nn\Lambda$ system such as the three-body binding energy.


\section{Amplitudes and coupled integral equations}
\label{sec:amp}

Throughout this study we employ the power counting rules suggested by Kaplan, Savage, 
and Wise (KSW) for the two-body sector~\cite{KSW98a,KSW98b}.
At LO we consider the $S$-wave two-body $nn$ and $n\Lambda$ interactions that leave 
us with the three channels in the two-body sector, namely, the $nn$ in ${}^1S_0$ state, 
$n\Lambda$ in ${}^1S_0$ state, and $n\Lambda$ in ${}^3S_1$ state. 
The KSW rules require the ``bubble'' diagrams in the two-body propagators of $nn$ and 
$n\Lambda$ to be resummed to infinite order, while the propagators are renormalized at LO 
using a single parameter, i.e., the respective $S$-wave scattering length.

For the three-body sector, we generally follow the prescription suggested by Bedaque, Hammer, 
and van~Kolck~\cite{BHV98}. 
Namely, a three-body contact interaction, which would otherwise be naively considered as a 
subleading contribution, should be promoted to LO.
This is because of the non-analytic ultra-violet (UV) enhancements from the cutoff dependence 
which appears whenever the three-body system exhibits the limit cycle behavior. 
Since there is no empirical information to fix the strength of the three-body interaction, however,
the three-body contact interaction is taken into account only when the limit-cycle behavior is 
explored, while we assume a vanishing binding energy of the $nn\Lambda$ system.%
\footnote{Therefore, this corresponds to the case when the energy of the system equals to its threshold 
energy.}
When we explore the $nn\Lambda$ binding energy, on the other hand, we exclude the contributions 
of the three-body contact term by focusing on the role of the two-body interactions.
In other words, instead of making a definite prediction on the binding energy of the $nn\Lambda$ 
system, we try to reveal some general features of the system.

\begin{figure*}[t]
\begin{center}
\includegraphics[width=1.5\columnwidth]{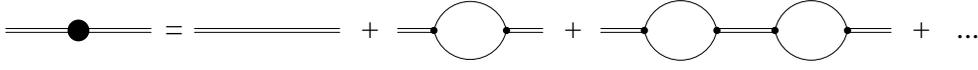}
\caption{
Diagrams for the dressed dibaryon propagator for the spin singlet $nn$ channel. 
The single and double lines denote the neutron and dibaryon fields, respectively.
}
\label{fig:dibaryon}
\end{center}
\end{figure*}

For the two-body sector, the Feynman diagrams for the dressed dibaryon field in the $nn ({}^1S_0)$ 
channel are shown in Fig.~\ref{fig:dibaryon}, which leads to the renormalized dressed dibaryon 
propagator at LO as
\begin{equation}
D_{s(nn)}(q_0^{},\bm{q}) = 
\frac{4\pi}{y_{s(nn)}^2 m_n^{} }
\frac{1}{
\frac{1}{a_{nn}^{}} - \sqrt{\frac14 \bm{q}^2- m_n^{} q_0^{}  - i \epsilon }- i \epsilon} ,
\label{eq:Dsnn}
\end{equation}
where $a_{nn}^{}$ is the scattering length of the neutron-neutron scattering in the $^1S_0$ channel, 
and $q_0^{}$ and $\bm{q}$ are generic off-shell energy and three-momentum, respectively. 
The loop diagrams are calculated using dimensional regularization with the power divergence
subtraction scheme~\cite{KSW98a,KSW98b} which introduces the subtraction scale parameter $\mu$.
The coupling constants in the Lagrangian are renormalized by using the $S$-wave effective range 
parameters, namely, the scattering length $a_{nn}^{}$ and the effective range $r_{nn}^{}$. 
Furthermore, the analyses in Refs.~\cite{BS00,AH04,AK06b} yield the sign factor
$\sigma_{s(nn)}^{}= -1$ and
\begin{eqnarray}
\frac{1}{a_{nn}^{}} = 
-\frac{4\pi\Delta_{s(nn)}}{m_n y_{s(nn)}^2} + \mu\,,
\qquad
r_{nn}^{} = \frac{8\pi}{m_n^2 y_{s(nn)}^2}\,.
\end{eqnarray}

\begin{figure*}[t]
\begin{center}
\includegraphics[width=1.5\columnwidth]{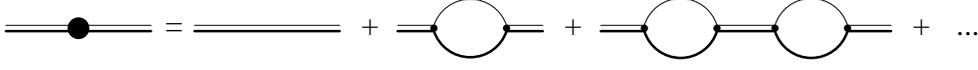}
\caption{
Diagrams for the dressed dibaryon propagator for the spin singlet and triplet $n\Lambda$ channels. 
The double line denotes the dibaryon field and the thin (thick) line the neutron 
($\Lambda$ hyperon) field. 
}
\label{fig:dibaryon2}
\end{center}
\end{figure*}

Similarly, for the $n\Lambda$ channels, Fig.~\ref{fig:dibaryon2} displays the Feynman diagrams 
for the dressed $n\Lambda$ propagators whose renormalized expressions are given by\,%
\footnote{The $i \epsilon$ prescription in the dressed propagators is 
understood in the same way as in Eq.~(\ref{eq:Dsnn}) for $D_{s(nn)}(q_0^{},\bm{q})$.}
\begin{widetext}
\begin{eqnarray}
D_{s(n\Lambda)}(q_0^{},\bm{q}) =
\frac{2\pi}{y_{s(n\Lambda)}^2 \mu_{(n\Lambda)}^{}}
\left[ \frac{1}{a_{s(n\Lambda)}^{}}
- \sqrt{- 2 \mu_{(n\Lambda)}^{} \left( q_0^{} - \frac{1}{2(m_n+m_\Lambda)} \bm{q}^2
\right) } \right]^{-1} ,
\label{eq:Ds}
\end{eqnarray}
for the spin singlet channel and
\begin{eqnarray}
D_{t(n\Lambda)}(q_0^{},\bm{q}) =
\frac{2\pi}{y_{t(n\Lambda)}^2 \mu_{(n\Lambda)}^{}}
\left[ \frac{1}{a_{t(n\Lambda)}^{}}
- \sqrt{ - 2\mu_{(n\Lambda)} \left( q_0^{} - \frac{1}{2(m_n+m_\Lambda)} \bm{q}^2
\right) } \right]^{-1} ,
\label{eq:Dt}
\end{eqnarray}
\end{widetext}
for the spin triplet channel, respectively, where $a_{s(n\Lambda)}^{}$ ($a_{t(n\Lambda)}$) 
is the $S$-wave scattering length of $n\Lambda$ scattering in the spin singlet (triplet) channel, 
and $\mu_{(n\Lambda)}$ is the reduced mass of the $n\Lambda$ system. 
Again the coupling constants of the two-body Lagrangian are renormalized by the $S$-wave 
effective range parameters and, as detailed in Refs.~\cite{BS00,AH04,AK06b}, we have 
$\sigma_{s,t(n\Lambda)}^{} = -1$ and 
\begin{eqnarray}
\frac{1}{a_{s,t(n\Lambda)}^{}} &=& 
-\frac{2\pi\Delta_{s,t(n\Lambda)}}{\mu_{(n\Lambda)}^{} y_{s,t(n\Lambda)}^2} + \mu \,
\nonumber \\
r_{s,t(n\Lambda)}^{} &=& \frac{2\pi}{\mu_{(n\Lambda)}^2 y_{s,t(n\Lambda)}^2}\, .
\end{eqnarray}

\begin{figure*}[t]
\begin{center}
\includegraphics[width=1.2\columnwidth]{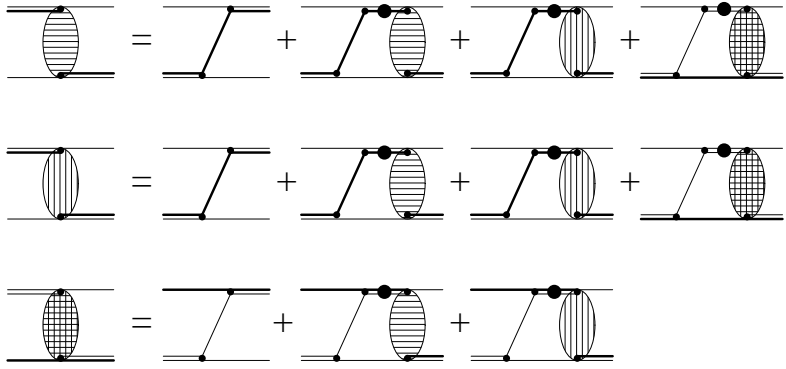}
\caption{
Diagrammatic representation of the coupled integral equations for the neutron and spin triplet 
$n\Lambda$ dibaryon elastic scattering in the spin-doublet channel without three-body contact 
interaction.
Blobs with the horizontal lines denote the elastic amplitudes for the neutron and spin triplet 
$n\Lambda$ dibaryon [$nt(n\Lambda)$] channel, while those with the vertical and crossed lines 
denote the two inelastic channels, namely, $nt(n\Lambda)$ to neutron and spin singlet 
$n\Lambda$ dibaryon, and $\Lambda$ hyperon and the spin singlet $nn$ dibaryon channels. 
See the captions of Figs.~\ref{fig:dibaryon} and \ref{fig:dibaryon2} as well.}
\label{fig:diagrams}
\end{center}
\end{figure*}

For the three-body part, we derive a set of coupled integral equations for the $S$-wave scattering 
of the neutron and the composite dibaryon triplet state $t(n\Lambda)$ in the momentum space. 
There are two possible allowed total-spin channels, namely, spin-3/2 and spin-1/2.
Since the integral equation for the spin-3/2 channel does not exhibit a limit cycle,%
\footnote{In the spin-3/2 channel, one obtains a single integral equation of which numerical solution 
does not yield the limit cycle behavior.}
we shall henceforth consider only the spin-1/2 channel as demonstrated by the diagrams in 
Fig.~\ref{fig:diagrams}. 
The coupled integral equations are expressed in terms of the three half-off-shell amplitudes, 
$a(p',p)$, $b(p',p)$, and $c(p',p)$, where $a(p',p)$ is the amplitude of elastic $nt(n\Lambda)$ 
scattering, $b(p',p)$ is that of inelastic $nt(n\Lambda)$ to $ns(n\Lambda)$ scattering, and $c(p'p)$ 
is that of inelastic $nt(n\Lambda)$ to $\Lambda s(nn)$ scattering, with $p'$ ($p$) being the 
relative off-shell (on-shell) momentum of the final (initial) two-body system. 
Explicitly, we have
\begin{widetext}
\begin{eqnarray}
a(p',p) &=& \frac12 m_\Lambda^{} y_{t(n\Lambda)}^{} K_{(a)}(E;p',p)
- \frac{1}{2\pi}\frac{m_\Lambda^{}}{\mu_{(n\Lambda)}^{}}
\int^{\Lambda_c}_0 dl \, l^2 K_{(a)}(E;p',l)
\frac{a(l,p)}{ \frac{1}{a_{t(n\Lambda)}^{}} 
- \sqrt{ \frac{\mu_{(n\Lambda)}^{}}{\mu_{n(n\Lambda)}^{}} l^2
 - 2 \mu_{(n\Lambda)}^{} E}}
\nonumber \\ && \mbox{}
- \frac{\sqrt3}{2\pi} \frac{m_\Lambda^{}}{\mu_{(n\Lambda)}^{}}
\int^{\Lambda_c}_0 dl \, l^2 K_{(a)}(E;p',l)
\frac{b(l,p)}{ \frac{1}{a_{s(n\Lambda)}^{}} -
 \sqrt{ \frac{\mu_{(n\Lambda)}^{}}{\mu_{n(n\Lambda)}^{}} l^2 
- 2 \mu_{(n\Lambda)}^{} E } }
\nonumber \\ && \mbox{}
- \frac{\sqrt6}{\pi} \int^{\Lambda_c}_0 dl \, l^2 K_{(b2)}(E;p',l)
\frac{c(l,p)}{ \frac{1}{a_{nn}^{}}
- \sqrt{ \frac{m_n^{}}{2 \mu_{(nn)\Lambda}^{}} l^2
- m_n^{} E } }  \,,
\label{eq:a}
\\
b(p',p) &=& \frac{\sqrt3}{2} m_\Lambda^{} y_{t(n\Lambda)} K_{(a)}(E;p',p)
- \frac{\sqrt3}{2\pi} \frac{m_\Lambda^{}}{\mu_{(n\Lambda)}^{}}
\int^{\Lambda_c}_0 dl \, l^2 K_{(a)}(E;p',l)
\frac{a(l,p)}{ \frac{1}{a_{t(n\Lambda)}^{}}
- \sqrt{ \frac{\mu_{(n\Lambda)}^{}}{\mu_{n(n\Lambda)}^{}} l^2
- 2 \mu_{(n\Lambda)}^{} E } }
\nonumber \\ && \mbox{}
+ \frac{1}{2\pi} \frac{m_\Lambda^{}}{\mu_{(n\Lambda)}^{}}
\int^{\Lambda_c}_0 dl \, l^2 K_{(a)}(E;p',l)
\frac{b(l,p)}{ \frac{1}{a_{s(n\Lambda)}^{}}
- \sqrt{ \frac{\mu_{(n\Lambda)}^{}}{\mu_{n(n\Lambda)}^{}} l^2
- 2 \mu_{(n\Lambda)}^{} E } }
\nonumber \\ && \mbox{}
+ \frac{\sqrt2}{\pi} \int^{\Lambda_c}_0 dl \, l^2 K_{(b2)}(E;p',l) 
\frac{c(l,p)}{ \frac{1}{a_{nn}^{}}
-\sqrt{ \frac{m_n^{}}{2\mu_{(nn)\Lambda}^{}} l^2
- m_n^{} E } }\, ,
\label{eq:b}
\\
c(p',p) &=& \sqrt{\frac32} m_n^{} y_{t(n\Lambda)} K_{(b1)}(E;p',p)
- \sqrt{\frac32} \frac{1}{\pi} \frac{m_n^{}}{\mu_{(n\Lambda)}^{}}
\int^{\Lambda_c}_0 dl \, l^2 K_{(b1)}(E;p',l)
\frac{a(l,p)}{ \frac{1}{a_{t(n\Lambda)}^{}}
- \sqrt{ \frac{\mu_{(n\Lambda)}^{}}{\mu_{n(n\Lambda)}^{}} l^2
- 2 \mu_{(n\Lambda)}^{} E } }
\nonumber \\ && \mbox{}
+ \frac{1}{\sqrt2\pi} \frac{m_n^{}}{\mu_{(n\Lambda)}^{}}
\int^{\Lambda_c}_0 dl \, l^2 K_{(b1)}(E;p',l)
\frac{b(l,p)}{ \frac{1}{a_{s(n\Lambda)}^{}}
- \sqrt{ \frac{\mu_{(n\Lambda)}^{}}{\mu_{n(n\Lambda)}^{}} l^2
- 2 \mu_{(n\Lambda)}^{} E } }  
\,,
\label{eq:c}
\end{eqnarray}
where $K_{(a)}(E;p',p)$ is the one-$\Lambda$-exchange interaction kernel, and 
$K_{(b1)}(E;p',p)$ and $K_{(b2)}(E;p',p)$ are the two possible one-neutron-exchange 
interaction kernels with $E$ being the total center-of-momentum energy of the three-body system. 
These interaction kernels are written as
\begin{eqnarray}
K_{(a)}(E;p',p) &=& \frac{1}{2p' p} \ln \left(
\frac{p'^2 + p^2 + \frac{2\mu_{(n\Lambda)}^{}}{m_\Lambda^{}} p'p - 2 \mu_{(n\Lambda)}^{} E}
   {p'^2 + p^2 - \frac{2\mu_{(n\Lambda)}^{}}{m_\Lambda^{}} p'p - 2 \mu_{(n\Lambda)}^{} E}
\right) ,
\\
K_{(b1)}(E;p',p) &=& \frac{1}{2p' p} \ln \left(
\frac{\frac{m_n^{}}{2\mu_{(n\Lambda)}^{}} p'^2 + p^2 + p' p - m_n^{} E}
   {\frac{m_n^{}}{2\mu_{(n\Lambda)}^{}} p'^2 + p^2 - p' p - m_n^{} E}
\right)  ,
\\
K_{(b2)}(E;p',p) &=& \frac{1}{2p' p}\ln\left(
\frac{p'^2 + \frac{m_n^{}}{2\mu_{(n\Lambda)}^{}} p^2 + p' p - m_n^{} E}
   {p'^2 + \frac{m_n^{}}{2\mu_{(n\Lambda)}^{}} p^2 - p' p - m_n^{} E}
\right)  ,
\end{eqnarray}
\end{widetext}
where $\mu_{n(n\Lambda)}$ and $\mu_{(nn)\Lambda}$ are reduced masses defined as
$\mu_{n(n\Lambda)}=  m_n^{} (m_n^{} + m_\Lambda^{})/(2m_n^{} + m_\Lambda^{})$
and $\mu_{(nn)\Lambda} = 2 m_n^{} m_\Lambda^{} / (2m_n^{} + m_\Lambda^{})$.
As mentioned earlier, the sharp momentum cutoff $\Lambda_c$ is introduced in the above integral 
equations assuming $E,p \sim 1/a_{nn}^{}, 1/a_{s,t(n\Lambda)}^{} \lesssim p' \ll \Lambda_c$.

While the inhomogeneous parts of the integral equations only set the overall low-energy scale of the 
problem in the asymptotic limit of the off-shell momenta $p'$ and $l$, i.e., $p',l \sim \Lambda_c$, 
the behavior of the solutions are completely determined by the homogeneous parts. 
This, however, leads to ambiguities in the solution when the equations are numerically solved 
in the UV limit ($\Lambda_c \to \infty$) without the three-body contact interaction. 
As  evident in the analysis below, we need to introduce a three-body counter term only for one of the 
three elastic scattering modes in the diagonal basis associated with the asymptotic limit cycle behavior. 
Nevertheless, when the integral equations of Eqs.~(\ref{eq:a})--(\ref{eq:c}) are numerically solved in the 
spin-isospin basis for intermediate momenta, one requires three-body contact interaction terms
with several different unknown couplings. 
In this preliminary analysis, this is represented by Eq.~(\ref{eq;3bci}), and we only need to modify 
the one-$\Lambda$-exchange kernel in the first two terms in Eq.~(\ref{eq:a}) as  
\begin{equation}
K_{(a)}(E;p',l) \to K_{(a)}(E;p',l) - \frac{g(\Lambda_c)}{\Lambda_c^2} .
\end{equation}

Before discussing the numerical solutions of the coupled integral equations of
Eqs.~(\ref{eq:a})--(\ref{eq:c}), let us consider the approximation suggested in Ref.~\cite{Hammer01}
for investigating the asymptotic nature of the solutions excluding the three-body interaction. 
Namely, since the mass difference between the $\Lambda$ hyperon and neutron is small, i.e., 
$\delta = (m_\Lambda^{} - m_n^{})/(m_\Lambda^{} + m_n^{}) \sim 0.1$, the corrections due to small 
$\delta$ to the integral equations may be ignored in the asymptotic limit.
Then we have
\begin{widetext}
\begin{equation}
\left( \begin{array}{c}
a(p') \cr
b(p') \cr
c(p') 
\end{array} \right) \approx \frac1\pi\int_0^\infty dl \, l \, \tilde{K}(p',l) 
\left(
\begin{array}{ccc}
\frac{2}{\sqrt3} & 2 & 2 \sqrt2 \cr
2 & - \frac{2}{\sqrt3} & - 2 \sqrt{\frac23} \cr
2\sqrt2 & - 2\sqrt{\frac23} & 0 
\end{array} \right)
\left( \begin{array}{c}
a(l) \cr
b(l) \cr
c(l) 
\end{array} \right) ,
\label{eq:M}
\end{equation}
\end{widetext}
where 
\begin{equation}
\tilde{K}(p',l) = \frac{1}{2p'l} \ln \left(
\frac{p'^2 + l^2 + p'l}{p'^2 + l^2 - p'l} \right) .
\label{eq:Ktilde}
\end{equation}
In the asymptotic limit of the respective amplitudes, the dependence on the on-shell momentum 
$p$ is implicitly understood. 
The integral equations can then be diagonalized to obtain generic homogeneous eigenvalue 
equations as
\begin{equation}
A_n(p') 
= \frac{\lambda_n}{2\pi} 
\int^\infty_0
\frac{dl}{p'}
\ln\left(
\frac{p'^2 + l^2 + p'l}
     {p'^2 + l^2 - p'l}
\right) A_n(l)\,,
\label{eq:An}
\end{equation}
where 
$\lambda_n$ with $n=1,2,3$ are the eigenvalues of the above $3 \times 3$ matrix obtained as
\begin{equation}
\lambda_n = \frac{8}{\sqrt3} \cos \left(\frac{2^n}{9}\pi \right) \,.
\label{eq:lambdan}
\end{equation}
Numerically, we have $\lambda_1 \approx {3.54}$, $\lambda_2 \approx 0.80$, and
$\lambda_3 \approx - 4.34$. 
The amplitudes $A_n(p',p)$ are given by
\begin{eqnarray}
A_n(p') &=&  
-\sqrt3\lambda_n\left(\frac{4}{\sqrt3}-\lambda_n\right) a(p')
\nonumber \\ &&
+\left(\frac{4}{\sqrt3}-\lambda_n\right)
\left(\frac{8}{\sqrt3}-\lambda_n\right) b(p')
\nonumber \\ &&
+\sqrt2\lambda_n\left(\frac{8}{\sqrt3}-\lambda_n\right)c(p')
\end{eqnarray}
with an arbitrary normalization.

The expression in Eq.~(\ref{eq:An}) has no scale dependence.
The scale invariance in the asymptotic limit, $p',l \sim \Lambda_c \gg p$, suggests that the 
$S$-wave projected amplitudes $A_n(p')$ must exhibit a power-law behavior as~\cite{BHV98}
\begin{equation}
A_n(p')\sim p'^{s-1}\,. 
\end{equation}
Then, by the Mellin transformation, Eq.~(\ref{eq:An}) becomes 
\begin{eqnarray}
1 &=& \frac{\lambda_n}{2\pi}\int^\infty_0 dx \ln\left(
\frac{x^2+x+1}{x^2-x+1}
\right)x^{s-1} 
\nonumber \\
&=& \frac{\lambda_n}{s}\frac{\sin\left(\frac{\pi}{6}s\right)}
{\cos\left(\frac{\pi}{2}s\right)} \, ,
\label{eq:Mellin}
\end{eqnarray}
where the solutions, in general, are complex-number functions and always come in pairs 
due to the additional inversion symmetry of $A(p')\to A(1/p')$ in
Eq.~(\ref{eq:An}), which is equivalent to $x\to 1/x$ in Eq.~(\ref{eq:Mellin}). 
Thus, there can be up to a quadruplet set of solutions, i.e., $\{\pm s,\pm s^*\}$, to the algebraic 
part of Eq.~(\ref{eq:Mellin}), all of which may not solve the integral equation. 
Of these, the physically acceptable solutions correspond only to those amplitudes which converge 
as $p'\to \infty$ and for which the Mellin transformation exists~\cite{Griess04,Griess05}. 
This scale-invariant equation then gives the condition to examine the limit cycle behavior of the 
spin-doublet $nn\Lambda$ system in the asymptotic limit. 
It is found that whenever the value of $\lambda_n$ exceeds the critical value $\lambda_c$, i.e., 
when $\lambda_n > \lambda_c = 6/\pi \approx 1.91$~\cite{BHV98}, the solutions become imaginary 
and a limit cycle appears in the system. 
Thus, only the first mode with $\lambda_1 \approx 3.54$ should correspond to a limit cycle behavior 
with an imaginary solution, $s = i s_0^{}$ with $s_0^{} = 0.80339 \cdots$\,.
In addition, for the purpose of renormalization it may be convenient to project out this mode from 
the other diagonal modes that do not exhibit the limit cycle.
To this end, the expression of the counter term in the limit, where the mass difference  between 
neutron and $\Lambda$ hyperon is ignored, 
can be constructed by considering the projection operator for the first mode, i.e., 
$(M-\lambda_2I)(M-\lambda_3I)$, where $M$ is the $3 \times 3$ matrix appearing in 
Eq.~(\ref{eq:M}) and $I$ is the $3 \times 3$ unit matrix.

In the following section we investigate the numerical solution of the coupled integral equations of 
Eqs.~(\ref{eq:a})--(\ref{eq:c}) valid for the non-asymptotic momentum range 
$p \lesssim p'\ll \Lambda_c$ with the physical $\Lambda$-$n$ mass difference. 
Our analysis will reveal that the characteristic cyclic behavior reminiscent of the asymptotic limit 
cycle clearly survives at low and intermediate momentum range. 
The solutions exhibit the strong cutoff dependence and are very likely sensitive to the 
short-distance dynamics. 
This is similar to what was seen, e.g., in the case of the triton~\cite{BHV99} as well as the 
hypertriton~\cite{Hammer01} channels. 
In contrast, the other two modes have well-behaved UV stable solutions that do not need 
a leading-order three-body renormalization. 
These modes do not exhibit the limit cycle and insensitive to the short-distance physics as seen, 
e.g., in the case of the spin quartet channel of $nd$ scattering~\cite{BV98a,Ando13}.


\section{Numerical results}
\label{sec:results}

In this section, we present our numerical results and investigate the singularities of the coupled 
integral equations for the $nn\Lambda$ system.
Our EFT has four parameters at LO to be determined in a phenomenological way.
They are the $nn$ and $n\Lambda$ $S$-wave scattering lengths, namely, $a_{nn}^{}$, 
$a_{s(n\Lambda)}^{}$, and $a_{t(n\Lambda)}^{}$, and the three-body coupling $g(\Lambda_c)$ 
with the cutoff $\Lambda_c$. 
In the present study, we determine these parameters as follows. 
First, we use the standard experimentally extracted value of the $nn$ scattering length,
$a_{nn}^{} \approx -18.5~\text{fm}$~\cite{DG87}. 
Since there is no empirical data for $n\Lambda$ scattering, however, we quote the theoretically 
calculated values of $^1a_{p\Lambda}$ and $^3a_{p\Lambda}$ reported in Ref.~\cite{HPKMNW13}, 
which are obtained based on the potential constructed in the chiral effective theory 
up to next-to-leading order (NLO).
Then, invoking isospin symmetry leads to the estimates: 
$a_{s(n\Lambda)}^{} \simeq -2.90~\text{fm}$ and $a_{t(n\Lambda)}^{} \simeq -1.60~\text{fm}$. 
Finally, there is no experimental or empirical information to fix the strength of the contact interaction 
$g(\Lambda_c)$.
Thus, we investigate the limit cycle behavior tuning the energy of the $nn\Lambda$ system to the 
threshold value, i.e., to vanishing three-body binding energy $B = 0$.

\begin{figure}[t]
\begin{center}
\includegraphics[width=\columnwidth]{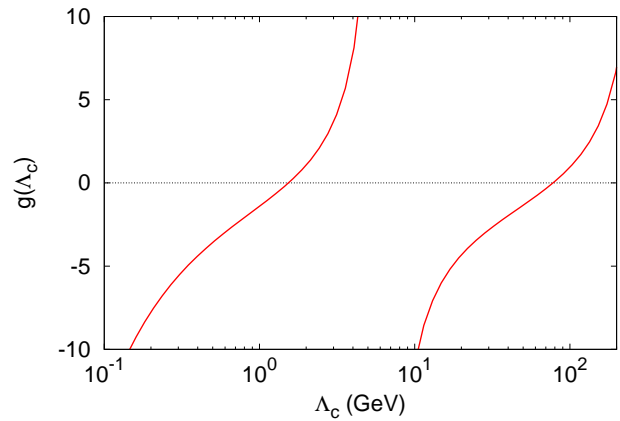}
\caption{(Color online)
Strength of the coupling $g(\Lambda_c)$ of the three-body contact interaction as a function of 
the cutoff $\Lambda_c$, where zero binding energy ($B=0$) of the three-body $nn\Lambda$ 
system is maintained.
}
\label{fig:gvsLam}
\end{center}
\end{figure}

In Fig.~\ref{fig:gvsLam}, we plot the strength of the three-body contact interaction $g(\Lambda_c)$ 
as a function of the cutoff $\Lambda_c$, which gives zero binding energy of the three-body system. 
The periodic RG evolution of $g(\Lambda_c)$ clearly indicates that the short-range part of the 
one-baryon-exchange interactions become singular, and thus the contact interaction is needed at \
LO for the renormalization of the singularities.
In addition, one finds that the interactions without the three-body force become stronger for larger 
values of the cutoff $\Lambda_c$. 
However, in the region of small $\Lambda_c \sim \Lambda_H$, the two-body interactions alone 
cannot generate attractive forces strong enough to form a bound state. 
This is manifest through the large negative values of $g(\Lambda_c)$ required to supplement the 
deficit in attractive forces for bound state formation. 
The lowest critical value of $\Lambda_c$, where $g(\Lambda_c)$ changes its sign in the above 
scenario, is found to be at around $1.5~\text{GeV}$. 
Consequently, if $\Lambda_c \gtrsim 1.5~\text{GeV}$, an $nn\Lambda$ bound state can be 
generated from two-body dynamics alone without the necessity of introducing three-body terms. 
On further increasing $\Lambda_c$, the second critical cutoff appears at 
$\Lambda_c \sim 80~\text{GeV}$, which indicates the onset of a second bound state.

Based on the \textit{universality\/} of the Efimov-like states, the occurrence of periodic critical 
points is expected as $g(\Lambda_{n+1}) = g(\Lambda_1) = 0$ for 
$\Lambda_{n+1} = \Lambda_1 \exp \left(n\pi/s_0 \right)$.
The result shown in Fig.~\ref{fig:gvsLam} reveals that $\Lambda_1 \simeq 1.54~\text{GeV}$ and 
$\Lambda_2 \simeq 77.8~\text{GeV}$, which leads to 
$s_0^{} = \pi/\ln(\Lambda_2/\Lambda_1) \approx 0.801$. 
This is surprisingly in good agreement with the value of $s_0^{}$ obtained in the previous 
section, suggesting an inherent universal behavior of the $nn\Lambda$ system governed 
by the asymptotic limit cycle behavior that is not `washed away' even for non-asymptotic 
momenta without considering the degenerate mass approximation.

\begin{figure}[t]
\begin{center}
\includegraphics[width=\columnwidth]{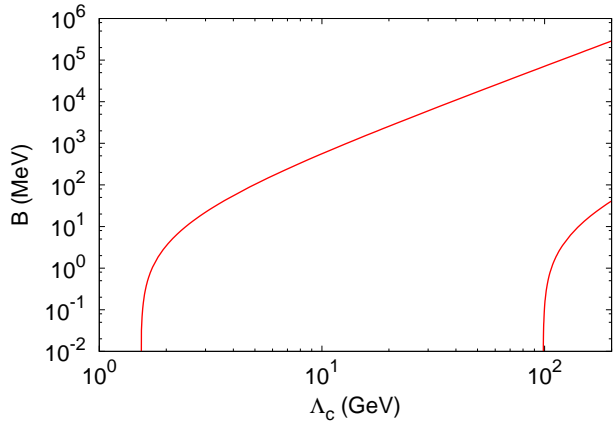}
\caption{(Color online)
Binding energy of the $nn\Lambda$ three-body system as a function of
the cutoff $\Lambda_c$ without the three-body contact interaction.
}
\label{fig:BvsLam}
\end{center}
\end{figure}

We now turn to the cutoff-dependence of the binding energy of the $nn\Lambda$ system
without the contact term, i.e., by setting $g(\Lambda_c) = 0$.
In Fig.~\ref{fig:BvsLam} we display the result for the the three-body binding energy $B$ without 
including the contact interaction. 
Clearly, the first three-body bound state appears at 
$\Lambda_c \sim 1.5~\mbox{GeV}$, and
the second one at $\Lambda_c \sim 80~\mbox{GeV}$.
This shows the periodic nature of the binding energies as a function of $\Lambda_c$. 
As seen in the figure, the binding energies of the shallow bound states, as they are formed, 
progressively increase from zero with increasing $\Lambda_c$, and eventually becoming very large, 
associated with deeply bound states.


\section{Summary and Discussion}
\label{sec:dis}

In the present study we investigated the possibility of a bound $nn\Lambda$ system in the 
pionless EFT at LO by exploring the structure of the coupled integral equations describing 
the system. 
Because of the limited information to fix the parameters of EFT, we are unable to arrive at a 
definitive conclusion on the formation of a bound $nn\Lambda$ state. However, by numerically 
solving the coupled integral equations including the three-body interaction with the binding 
energy fixed to a certain three-body threshold value, e.g., $B=0$ as used in this study, the 
coupling $g(\Lambda_c)$ was found to develop cyclic singularities, which is a characteristic of 
Efimov-like bound states. 
In addition, we studied the role of short range two-body mechanisms involving the one-$\Lambda$ 
and one-neutron exchange interactions by setting $g(\Lambda_c) = 0$.
Our analysis reveals that bound states appear without the requirement of the contact 
interaction only if the cutoff was chosen unnaturally large, 
$\Lambda_c \gtrsim 1.5~\mbox{GeV} \gg \Lambda_H$,
which indicates the nontrivial role of short-range two-body mechanisms that are beyond the scope
of this work.

As mentioned before, the predictability of our approach relies on the knowledge of the values of 
four low-energy parameters. 
In the present study, we use the empirically known value of the $nn$ scattering length, i.e.,
$a_{nn}^{} \approx -18.5$~fm~\cite{DG87}.  
Since there is no data for the $n\Lambda$ scattering lengths, we invoke isospin symmetry to 
determine their values from the $p\Lambda$ scattering lengths.
However, the poor statistics of experimental data do not allow precise partial wave analyses and 
the estimated scattering lengths have inevitably large uncertainties. 
For example, in Ref.~\cite{CJF96}, the ranges of the $p\Lambda$ scattering lengths are given as 
$a_{s(n\Lambda)} \simeq (-1.85 \sim -2.78)~\mbox{fm}$ and 
$a_{t(n\Lambda)}^{} \simeq (-1.04 \sim -1.90)~\mbox{fm}$ depending on the model potential.  
In the present study we used the NLO chiral EFT results reported in Ref.~\cite{HPKMNW13}:
$a_{s(n\Lambda)}= -2.90~\mbox{fm}$ and $a_{t(n\Lambda)}=-1.60~\mbox{fm}$.
However, the previous version of the chiral EFT calculation at LO reported significantly small values 
for the scattering lengths,
$a_{s(n\Lambda)}^{} \simeq - 1.91~\mbox{fm}$ and 
$a_{t(n\Lambda)}^{} \simeq -1.23~\mbox{fm}$~\cite{PHM06}.
(See also Ref.~\cite{Nogga13}.)
Thus, the currently estimated values of the $p\Lambda$ scattering lengths 
are beset with certain uncertainties which evidently propagate into our analysis.%
\footnote{From Refs.~\cite{HPKMNW13,Nogga13}, we find that the $N\Lambda$ 
scattering lengths converge to values around $-2.9$ to $-2.5$~fm for the singlet case and 
$-1.7$ to $-1.4$~fm for the triplet case.}
For instance, if we use the values of Ref.~\cite{PHM06}, the first bound state of the system 
appears at $\Lambda_c \simeq 2~\mbox{GeV}$, which should be compared with 
$\Lambda_c \simeq 1.5~\mbox{GeV}$ obtained with the values of Ref.~\cite{HPKMNW13}. 
However, this does not have any significant effect on the cyclic nature of the bound system.

In contrast with the case of the scattering lengths, there is no \textit{a priori} way to determine 
the strength of the three-body contact interaction without an essential information on a measured 
three-body observable. 
Thus, owing to the absence of enough information to fix the parameters of the EFT developed in
this work, our current results are not conclusive on the existence of the putative $nn\Lambda$ 
bound state.
Since we treat the effective $N\Lambda$ interaction as point-like, short range mechanisms such 
as the two-pion exchange interactions appearing as box diagrams with $N\Lambda$-$N\Sigma$ 
mixing, or other meson exchanges such as the $\eta$ and $K$ are to be implicitly taken into 
account by the contact interactions. 
Thus, without a three-body contact interaction, a reasonable value of the cutoff would be 
commensurate with the energy scale of the two-pion/$\eta$-meson/$K$-meson exchanges 
of the $N\Lambda$ interaction, which leads to $\Lambda_c \sim 300 - 500$~MeV.
This value is consistent with the values of $\Lambda_c$ found in similar context in the literature. 
For example, when the triton and \nuclide[6][\Lambda\Lambda]{He} are treated as  
three-nucleon and $\Lambda\Lambda\alpha$ bound systems, respectively, the cutoff values
are obtained as $\Lambda_c \simeq 380$~MeV for the case of the triton~\cite{AB10b}, and 
$\Lambda_c \simeq 400 - 570$~MeV for the case of \nuclide[6][\Lambda\Lambda]{He}~\cite{AO14}, 
depending on the adopted value of the $S$-wave $\Lambda\Lambda$ scattering length 
with the three-body contact interactions turned off. 
This contrasts with our case as we indeed need a very large cutoff to a form bound state without 
the three-body contact term. 
Therefore, the short-range two-body attractive mechanisms discussed above are unlikely to 
generate enough attractions in the one-$\Lambda$-exchange or one-neutron-exchange interaction 
channels that could lead to $nn\Lambda$ bound states. 
This ultimately would make it difficult to form an $nn\Lambda$ state in nature. 
However, further rigorous investigations, both theoretical and experimental, are necessary to pin 
down to true character of the $nn\Lambda$ system.


\acknowledgments

S.-I.A. is grateful to the organizers of the program ``Bound states and resonances in effective 
field theories and lattice QCD calculations'' in Centro de Ciencias de Benasque Pedro Pascual 
for hospitality and supports. 
U.R. acknowledges Sunmoon University for the warm hospitality and financial support for his stay 
during the initial stage of this work.  
S.-I.A. and Y.O. were supported by the Basic Science Research Program through the National 
Research Foundation of Korea funded by the Ministry of Education of Korea under Grant Nos.\ 
NRF-2012R1A1A2009430 and NRF-2013R1A1A2A10007294, respectively.
This work was also supported in part by the Ministry of Science, ICT, and Future Planning (MSIP) 
and the National Research Foundation of Korea under Grant No.\ NRF-2013K1A3A7A06056592 
(Center for Korean J-PARC Users).

\end{document}